\documentclass[aps,prl,twocolumn,groupedaddress,showpacs]{revtex4-1}

\usepackage{stmaryrd}
\usepackage{siunitx}
\usepackage{mathtools}
\usepackage{microtype}
\usepackage{graphics}

\DeclarePairedDelimiter\abs{\lvert}{\rvert}
\DeclarePairedDelimiter\avg{\langle}{\rangle}
\newcommand{\qavg}[1]{\overline{#1}}
\newcommand{\ddf}{\,\mathrm{d}}
\newcommand{\df}{\mathrm{d}}

\renewcommand{\v}[1]{\mathbf{#1}}

\begin{document}

\title{Persistence-Length Renormalization of Polymers in a Crowded Environment of Hard Disks}


\author{S. Sch\"obl}
\author{S. Sturm}
\author{W. Janke}
\email[]{wolfhard.janke@itp.uni-leipzig.de}
\author{K. Kroy}
\email[]{klaus.kroy@uni-leipzig.de}
\affiliation{Universit\"at Leipzig,
  Institut f\"ur Theoretische Physik,
  Postfach 100\,920,
  D-04009 Leipzig, Germany}


\date{\today}

\begin{abstract}
  The most conspicuous property of a semiflexible polymer is its persistence length, defined as the decay length of tangent correlations along its contour. 
  Using an efficient stochastic growth algorithm to sample polymers embedded in a quenched two-dimensional hard-disk fluid, we find apparent wormlike chain statistics with a renormalized persistence length.
  We identify a universal form of the disorder renormalization that suggests itself as a quantitative measure of molecular crowding.
\end{abstract}

\pacs{87.15.A-, 87.15.ak, 87.16.Ln}

\maketitle

Single-molecule experiments have established the wormlike chain (WLC) as a standard model for semiflexible biopolymers~\cite{bustamante2003, rief1997, gittes1993, marko1995stretching, natcomm2013}.
It emerges from the classical Heisenberg model for a ferromagnetic spin chain in the continuum limit, where the spins merge into an inextensible space curve \(\v r_s\) and the exchange interaction between adjacent spins (or unit tangent vectors \(\v r_s'\)) turns into an energetic penalty for bending.
The resulting continuum Hamiltonian \(\frac\kappa2\int {(\v r_s'')}^2 \df s\) is also familiar from continuum mechanics as a model for a slender rod with bending rigidity \(\kappa\)~\cite{landau1986elasticity}.
The WLC model accurately describes a very diverse range of polymers, including DNA~\cite{bustamante2003}, muscle protein~\cite{rief1997}, filamentous actin~\cite{gittes1993}, as well as synthetic carbon nanotubes~\cite{sano2001, Barnard2012}.

Exploiting the analogy of the WLC with the Heisenberg magnet, it can be shown~\cite{thompson1988} that for an isolated WLC in \(d\) dimensions, the equilibrium tangent-tangent correlation function decays exponentially,
\begin{equation}
  \label{eq:ellp}
  \langle \v r'_s\cdot \v r'_{s+\Delta s} \rangle = \exp(-|\Delta s|/\ell_p)\,,
\end{equation}
and thus defines a \emph{thermal} persistence length
\begin{equation}\label{eq:ellp-thermal}
  \ell_p = \frac{2\kappa}{(d - 1) k_B T}.
\end{equation}

The analytically solvable case of a single polymer in isolation maps well to single-molecule experiments but bears little resemblance to the disordered environment provided by a surrounding polymer network or even the cytoplasm~\cite{ellis2001, weiss2004anomalous,
  hofling2013anomalous}.
Although much work has been done on both local~\cite{hinsch2007, glaser2010, glaser2011, sussman2011, hinsch2009} and global~\cite{cates-ball1988, dua-vilgis2004} properties of semiflexible polymers in disorder, it remains unknown just how much of the WLC survives in the presence of a disordered environment.
The exponentially decaying tangent-tangent correlation function in Eq.~\eqref{eq:ellp} might for example turn into some nonexponential function of \(\Delta s\), or it might remain exponential and thus define a \emph{renormalized} persistence length, which then may or may not agree with its thermal value, given by Eq.~\eqref{eq:ellp-thermal}.
In the last case, one can imagine that the persistence length increases (due to channel formation~\cite{schoebl2012}) or decreases (due to crumpling induced by the obstacles) with respect to its thermal value.

For an isolated WLC, the exponential decay described by Eq.~\eqref{eq:ellp} derives from a random walk in tangent space, which is most easily visualized by imagining a thermal ensemble of polymers with one of their ends held clamped along a given direction.
Their free ends will then point along the same direction if the polymers are very short, but stray away from it diffusively if the polymers are longer.
Conversely, if the same ensemble is exposed to a quenched random background that is in some regions more favorable than in others, the free polymer ends will naturally gravitate towards the more favorable regions.
It has long been established that this can give rise to a superdiffusive growth of transverse fluctuations for directed polymers in random media~\cite{kardar2007fields, zhang1995}.
Recent results by Boltz and Kierfeld show that it can also give rise to a superdiffusive growth of tangent fluctuations for stiff semiflexible polymers exposed to \(\delta\)-correlated quenched Gaussian disorder~\cite{kierfeld2013}.
A superdiffusive growth of tangent fluctuations translates to a faster decay of tangent-tangent correlations compared to an isolated WLC, which on a scaling level can be captured by an ``effective disorder-induced persistence length''~\cite{kierfeld2013} that is smaller than the thermal value Eq.~\eqref{eq:ellp-thermal}.
It also translates, however, to a tangent-tangent correlation function decaying faster than exponentially, and thus does not define a persistence length in the strict sense of Eq.~\eqref{eq:ellp}.

To find out whether a renormalized persistence length in the strong sense of Eq.~\eqref{eq:ellp} might be observed under more realistic conditions, we have performed extensive numerical simulations of two-dimensional semiflexible polymers in a quenched equilibrium hard-disk fluid.
As we demonstrate below, this disorder-averaged ensemble of test polymers indeed exhibits a renormalized persistence length in the strong sense, which can account also for the dominant effect of the crowding onto the polymer end-to-end distribution.
This suggests that even if a shape analysis of microscopic {\em in vivo\/} imagery seems to agree perfectly well with standard WLC results, the persistence length inferred from the data may deviate significantly from the polymer's thermal persistence length that obeys Eq.~\eqref{eq:ellp-thermal}.
For sufficiently stiff polymers, we find that the renormalized persistence length can be determined uniquely from the thermal persistence length and an auxiliary quantity that characterizes the disordered environment.
This feature, which is known to hold true in the double asymptotic limit of high stiffness and \(\delta\)-correlated
Gaussian disorder~\cite{hwa1994, kierfeld2013}, but to our knowledge never has been observed in a more realistic setting, opens a novel way to employ polymers of known stiffness as quantitative probes of molecular crowding.

Our simulations are based on the discrete representation of the polymer, which
is constrained to lie in a plane (\(d=2\)).  The discrete (Heisenberg) and
continuum (WLC) forms of the Hamiltonian read
\begin{equation}
  \label{hamiltonian-2d}
  \mathcal{H} =  - \frac{\kappa}{b} \sum_{i=1}^{N-1} \v{t}_i \cdot
  \v{t}_{i+1}  \simeq \frac{\kappa}{2} \int_0^L {(\theta_s')}^2 \ddf s\,.
\end{equation}
The unit tangent \(\v{r}'_s\) at arclength position $s=ib$, corresponding to the
\(i'\)th spin \(\v{t}_i \equiv (\v{r}_{i+1} - \v{r}_i)/b\), has been identified
with its angle \(\theta_s\) in the plane.  It diffuses freely on the unit circle
as a function of the arclength, so that the increments \(\Delta \theta\) are
Gaussian distributed according to
\begin{equation}
  \label{eq:tangent-distribution}
  \frac1{\sqrt{2\pi}\sigma}
  \exp\left[-\frac{{(\Delta \theta)}^2}{2\sigma^2}\right]\,, \quad
  \sigma^2 \equiv \frac{k_B T}\kappa \Delta s\,.
\end{equation}
This induces an exponential decay of tangent correlations, which yields, by comparison with Eq.~\eqref{eq:ellp}, the thermal persistence length \(\ell_p = 2\kappa/k_B T\).
In the following, we subject the polymer to a
random, quenched and statistically isotropic background potential
\(V(\v{r})\). In contrast to earlier analytical works on the
subject~\cite{dua-vilgis2004, cates-ball1988} we are free to do away with the
simplifying assumptions of a vanishing correlation length or a Gaussian
distribution of potential energies; instead we adhere closely to the spirit of
molecular crowding by assuming sterically interacting obstacles that are small
compared to the polymer, but still large on the monomer scale. As a paradigmatic
representative of steric disorder, we consider an equilibrated (but quenched
with respect to the polymer) hard-disk fluid at several area filling fractions
\(\phi\), ranging from \(\SI{40}{\percent}\) to the verge of the freezing
transition at \(\phi \approx \SI{70}{\percent}\)~\cite{krauth2011}.  Then
\(V(\v{r})\) takes the values \(\infty\) or \(0\), depending on whether the
polymer penetrates any of the disks or not.
Since in an unbounded quenched system the polymer as a whole will gravitate towards ever more favorable regions~\cite{cates-ball1988}, thus in our case producing a trivial ensemble of straight rods in the limit of large persistence lengths and infinitely large systems, we eliminate the dependence on system volume by fixing one polymer end at the origin, \(\v{r}_0 = (0,0)\), which one might think of as a membrane-bound anchor point in the context of biopolymers in cells.

The extreme strength and density of environmental interactions present a
formidable challenge to conventional Monte Carlo simulation schemes, which we
found hard to overcome even using a sophisticated multicanonical histogram
reweighting procedure~\cite{schoebl2011}. We have therefore adopted a
breadth-first growth algorithm~\cite{orland1999} that resolves this difficulty
by {\em circumventing\/} energy barriers instead of trying to cross
them~\cite{schoebl2011, schoebl2012}.
For a given disorder realization, the algorithm starts with an ensemble of monomers fixed at the origin and then performs \(N - 1\) successive growth steps to extend each ensemble member to a polymer of the desired length.
During each growth step, every polymer in the ensemble branches out into a small set of new trial configurations through the addition of monomers pointing in random directions.
This incurs a severalfold increase of the ensemble population.
Next, the Boltzmann weight of each new trial configuration is calculated.
Each trial configuration is then either eliminated or replicated probabilistically, such that (i) the overall ensemble population is held approximately constant to prevent memory and processing time requirements from growing exponentially and (ii) the occurrence probability of any given configuration agrees with its equilibrium probability  \(\propto \exp(-\beta \mathcal{H}[\{\v r_i\}])\), thus establishing thermal equilibrium after each growth step.

\begin{figure*}
  \includegraphics{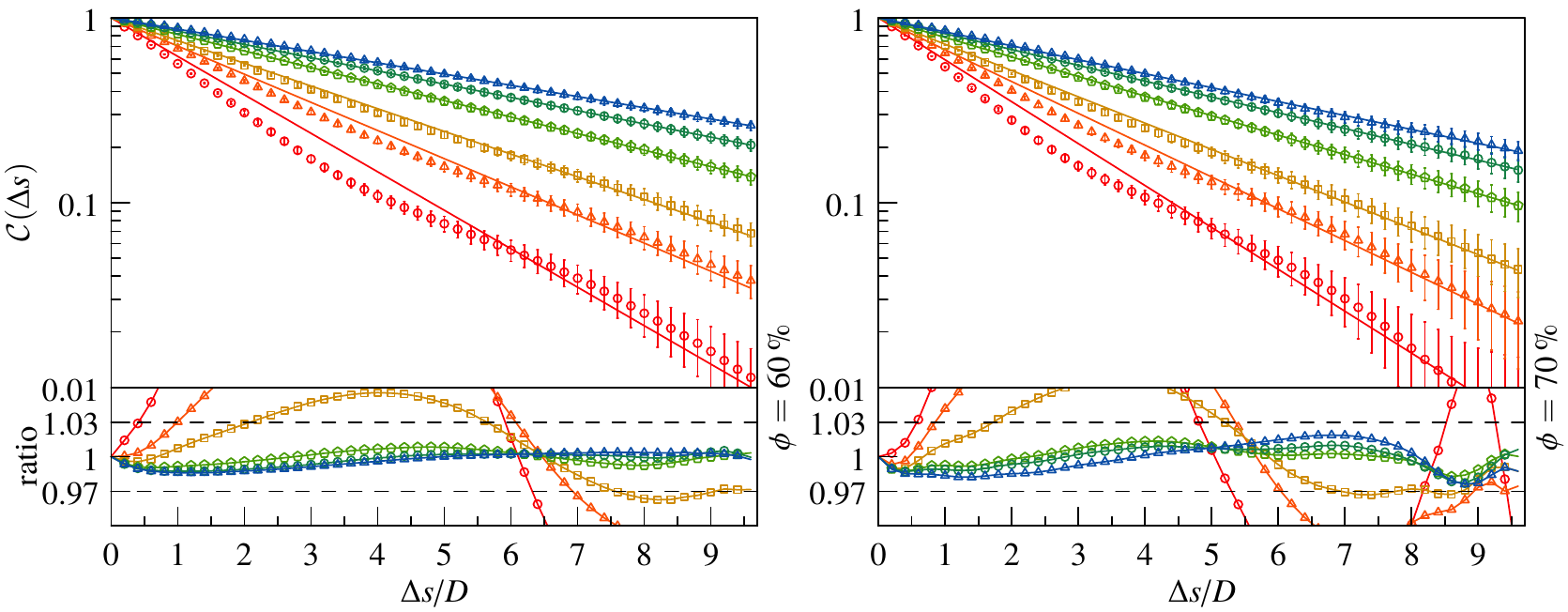}%
  \caption{\label{fig:tangent-correlations}Numerically obtained
    disorder-averaged tangent correlation function \(\mathcal{C}\) as a function
    of the reduced backbone distance for background filling fractions
    \(\phi=\SI{60}{\percent}, \SI{70}{\percent}\) and thermal persistence lengths
    \(\ell_p/D = 2, 3, 4, 6, 8, 10\) (from bottom to top). Solid lines indicate
    our exponential fits \(\mathcal{C}(\Delta s) = \exp(-\Delta
    s/\ell_p^\ast)\). For stiff polymers \(\ell_p \gtrsim 5D\) the relative
    error of the exponential fit remains bounded to about \(\SI{3}{\percent}\)
    (bottom panels).}
\end{figure*}

\begin{figure}
  \includegraphics{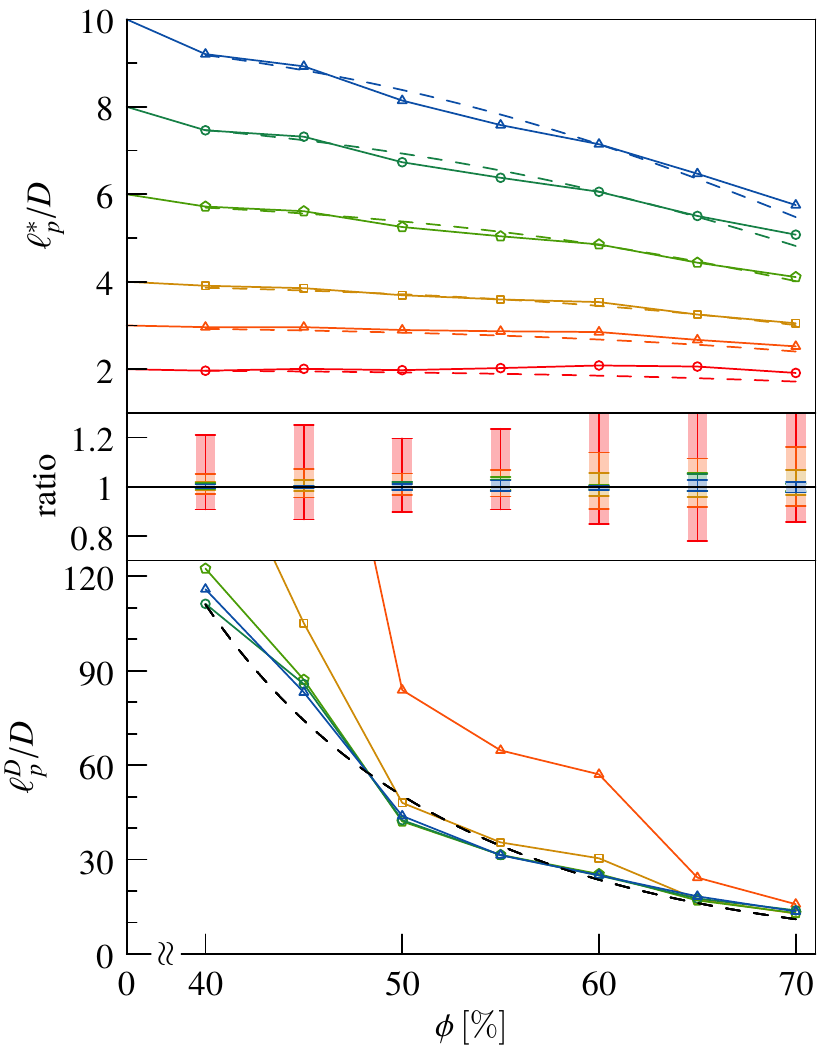}%
  \caption{\label{fig:renormalized-persistence-length} Renormalized persistence
    lengths \(\ell_p^\ast\) resulting from the exponential fits in
    Fig.~\ref{fig:tangent-correlations} for thermal persistence lengths
    \(\ell_p/D = 2, 3, 4, 6, 8, 10\) (upper panel/solid lines, from bottom to
    top) and Eq.~\eqref{eq:effective-lp} (dashed). The maxima of the relative
    fit errors (lower panel of Fig.~\ref{fig:tangent-correlations}) are
    indicated as bars (middle panel). For \(\ell_p > 2D\) the disorder
    persistence length \(\ell_p^D = (1/\ell_p^\ast - 1/\ell_p)^{-1}\) (bottom
    panel) quickly converges onto a single \(\ell_p\)-independent master curve
    as implied by Eq.~\eqref{eq:effective-lp-limit} (dashed line).}
\end{figure}

We analyze our numerical data first in terms of the length-averaged tangent correlation function
\begin{equation}\label{eq:quenchedtt}
  \mathcal{C}(\Delta s) \equiv \frac{1}{L - \Delta s}
  \int_0^{L - \Delta s} \!\!\!\qavg{\avg{\v{t}_{s}
      \cdot \v{t}_{s + \Delta s}}} \ddf s,
\end{equation}
where the overbar denotes the additional disorder average. As
demonstrated in Fig.~\ref{fig:tangent-correlations}, \(\mathcal{C}(\Delta s)\)
still decays exponentially in the mean.  Disorder-induced
non-exponential modulations are found to decay in amplitude as the thermal
persistence length \(\ell_p=2\kappa/(k_{B}T)\) increases compared to the
obstacle size \(D\) and the polymer length \(L\)---or, equivalently, upon
decreasing the temperature \(T\). Already at modest thermal persistence
lengths \(\ell_p \approx L/2\), a value easily realizable in experiment with
filamentous actin or carbon nanotubes, the deviations from perfect
exponentiality nowhere exceed \(\SI{3}{\percent}\).
This behavior, which we would not necessarily have anticipated, justifies the notion of a persistence-length renormalization~\footnote{That the effect of crowding on the statistical conformation of the test polymer can essentially be captured by a renormalized persistence length \(\ell_p \to \ell_p^\ast\) does not imply a renormalized bending rigidity \(\kappa \to \kappa^\ast\), as introduced in a different context~\cite{gutjahr2006}; \(\kappa\) is a microscopic parameter of the Hamiltonian in Eq.~\eqref{hamiltonian-2d}, \(\ell_p\) is a statistical observable defined by Eq.~\eqref{eq:ellp}.}.
Every disorder-averaged polymer ensemble defined by a thermal persistence length and a given density and size of background disks can therefore be characterized by an apparent renormalized persistence length \(\ell_p^\ast\) inferred from fitting \(\exp[-\Delta s/\ell_p^\ast]\) to Eq.~\eqref{eq:quenchedtt}, as exemplified in Fig.~\ref{fig:tangent-correlations}.

The fit results are shown in Fig.~\ref{fig:renormalized-persistence-length} for
two representative disorder filling fractions \(\phi\) and thermal persistence
lengths \(\ell_p\) between \(2D\) and \(10D\).  The total polymer length \(L\)
is \(10\) disk diameters \(D\) and the discretization length \(b = D/5\).
The inferred values for \(\ell_p^\ast\) systematically decrease with increasing disorder filling fraction for all but the smallest thermal persistence length \(\ell_p = 2\).
While for \(\delta\)-correlated Gaussian disorder, the renormalized persistence length \(\ell_p^\ast\) should actually change with \(L\)~\cite{kierfeld2013}, no significant $L$-dependence of \(\ell_p^\ast\) can be detected within the range of contour lengths accessible in our simulations~\footnote{Note that even within the strong-disorder regime as defined by Boltz and Kierfeld, the predicted deviation from normal tangent diffusion only scales with a small power of \(L\), \(\Delta \theta^2 \sim L^{1.18\ldots}\)~\cite{kierfeld2013}.
To clearly distinguish between a true exponential decay of tangent-tangent correlations that extends to \(L \to \infty\), and a slow crossover akin to Eq.~(43) in Ref.~\cite{kierfeld2013} would require us to further extend \(L\) beyond the limits of current numerical methods.
This does not diminish the relevance of our results to biophysical experiments, where polymer lengths typically do not vary over several orders of magnitude.}.
As \(L\) in our simulations is comparable to the thermal persistence length \(\ell_p\), and several times larger than the obstacle size, these contour lengths should closely approximate typical experimental conditions in cell biophysics.

The observed exponential scaling of tangent-tangent correlations implies that the renormalized persistence length \(\ell_p^\ast\) still derives from a diffusion process in tangent space (as for a free WLC), albeit with a renormalized diffusivity.
It is useful to note at this point that our model system exhibits, in the limit of vanishing obstacle size \(D\to 0\) and diverging persistence length \(\ell_p \to \infty\), the so-called {\em tilt symmetry\/}~\cite{hwa1994, kierfeld2013}.
It causes the disorder-averaged angular fluctuations to
separate into an unperturbed, ``thermal'' part and a disorder-induced
part. Together with the observed \(L\)-independent persistence length
renormalization, we thus find the asymptotic relation
\begin{equation}
  \label{eq:tangent-angles}
  \qavg{\avg{(\Delta \theta^2)}}/(2\Delta s) =  1/\ell_p^\ast 
  = 1/\ell_p + 1/\ell_p^D\,,
\end{equation}
which defines a ``disorder persistence length'' $\ell_p^D(\phi, D, \ell_p)$ that
should converge onto a \(\ell_p\)-independent master curve, in the limit
\(\ell_p \to \infty\).
This prediction is in fact well supported by our data, despite the finite size of the obstacles; see Fig.~\ref{fig:renormalized-persistence-length} (bottom).
As a consequence, we
can even rationalize the form of this master curve, on a scaling level. Namely,
the pure disorder effect onto the polymer conformation may simply be represented
as a succession of \(D\)-sized deflections, separated by some distance
\(\ell\) that roughly corresponds to the ``mean free path'' between subsequent
polymer-obstacle collisions. Each real-space deflection of size \(D\) amounts to
a rotation \(\delta\theta = D/\ell\) in tangent space, hence giving rise to a
disorder persistence length \(\ell_p^D = 2\ell^3/D^2\). The mean free path
\(\ell\) should scale with \(\smash{\phi^{-1/2}}\) at small disorder densities and approach a
value on the order of \(D\) at hexagonal close packing,
\(\ell(\phi_{\text{hcp}} \approx \SI{90.7}{\percent}) \approx D\). Indeed we
find that with the semi-empirical form \(\ell/D\approx 5.3 (\smash{\phi^{-1/2}} -
\smash{\phi_{\text{hcp}}^{-1/2}}) + 1 \) the argument provides an accurate analytical
parametrization of the renormalized persistence length
\(\ell_p^\ast(\phi,D,\ell_p)\),
\begin{equation}
  \label{eq:effective-lp}
  D/\ell_p^\ast \approx D/\ell_p +
  \frac{1}{2}{[5.3(\phi^{-1/2} - \phi_{\text{hcp}}^{-1/2}) + 1]}^{-3}\;.
\end{equation}
Figure~\ref{fig:renormalized-persistence-length} compares
Eq.~\eqref{eq:effective-lp} and its polymer-independent asymptotics for
\(\ell_p\to\infty\),
\begin{equation}
  \label{eq:effective-lp-limit}
  \ell_p^\ast \sim \ell_p^D(\phi, D) \approx 2D 
  {[5.3(\phi^{-1/2} - \phi_{\text{hcp}}^{-1/2}) + 1]}^3,
\end{equation}
to our numerical data. Note that both \(\smash{\ell_p^\ast}\) and \(\smash{\ell_p}\) can be
determined experimentally, so that \(\ell_p^D = \ell_p^\ast(\ell_p \gg D)\)
provides a practical quantitative measure of the environmental disorder
strength. Equation~\eqref{eq:effective-lp-limit} and possible refinements for
polydisperse obstacle sizes, might thus prove useful in future attempts to
quantify cellular crowding in terms of the density and size (distribution) of
the steric obstacles.

Although effective semiflexibility re-emerges on the global level of tangent
correlations, more localized observables must bear witness to the presence of
disorder correlations, allowing one to distinguish experimentally between
``true'' and ``renormalized'' semiflexibility, and providing further information
on the nature of the obstacles and their correlations. Here, we discuss the
disorder-averaged radial distribution function,
\begin{equation}
  \label{eq:quenchedrdf}
  \mathcal{P}(r) \equiv 2\pi r\qavg{\avg{\delta (\v r- \v r_L + \v r_0)}}\,,
  \qquad 
  \text{with }r\equiv |\v r|  \,,
\end{equation}
as an important example of such more local observables.
\begin{figure}
 \includegraphics{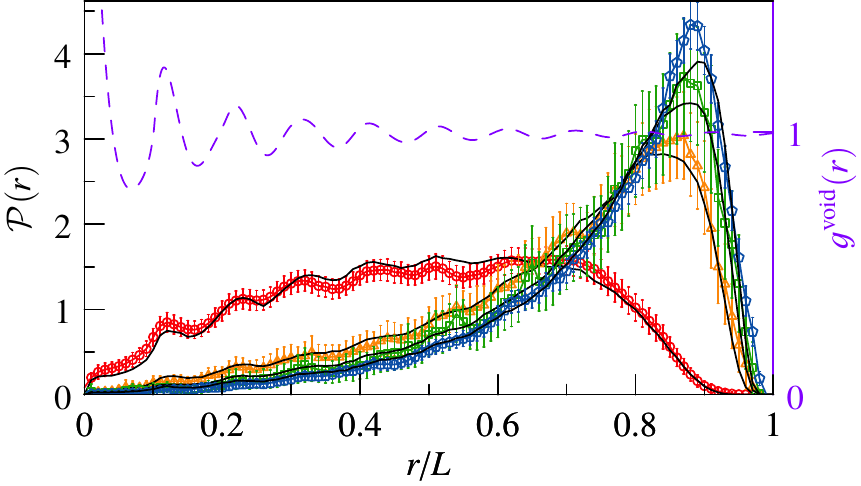}%
 \caption{\label{fig:radialdistributions}Numerical disorder-averaged radial
   distribution functions \(\mathcal{P}(r)\) for thermal persistence lengths
   \(\ell_p/D = 2, 6, 8, 10\) (data) and \(\phi=\SI{70}{\percent}\). Solid lines
   represent a parameter-free comparison with free WLC radial distribution
   functions evaluated for the corresponding renormalized persistence lengths
   \(\ell_p^\ast\) obtained from Fig.~\ref{fig:renormalized-persistence-length}
   and multiplied by the normalized void distribution function \(g^{\rm void}(r)\)
   (dashed).}
\end{figure}
As seen from Fig.~\ref{fig:radialdistributions}, \( {\cal P}(r)\) picks up the
fluid structure of the background, as the ensemble branches out inside the more
expansive voids and circumvents denser regions. To a first approximation,
\(\mathcal{P}(r)\) factorizes into two contributions: the free WLC radial
distribution~\cite{wilhelm1996radial, hamprecht2004end} with the renormalized
persistence length \(\ell_p^\ast\) and a factor weighing the relative abundance
of void space at a given distance \(r\). As demonstrated in
Fig.~\ref{fig:radialdistributions}, the latter is well represented by the ``void
space distribution function'',
\begin{equation}
  \label{eq:vsdf}
  g^{\rm void}(r) \propto
  r^{-1} \qavg{\int \ddf \v{r}'
      \delta(\abs{\v{r}'} - r) 
      e^{-\beta V(\v{0})} e^{-\beta V(\v{r}')}
    }\,,
\end{equation}
a close cousin to the radial distribution function known from liquid-state
theory.
  
Our results suggest that the molecular crowding in the cytoplasm of cells will
crumple embedded cytoskeletal polymers. Although we are not the first to predict
that a quenched disordered background should induce a renormalized persistence
length~\cite{kierfeld2013}, we were able to show explicitly that tangent
correlations remain exponential, even at the highest filling fractions and for
finite obstacle size. This indicates that the common practice of performing a
static shape analysis of single fluorescently labeled polymers {\em in vivo} or
in {\em in-vitro\/} reconstituted polymer solutions and networks requires special caution. It may not yield
a reliable estimator of intrinsic polymer stiffness, even if its results look
deceptively consistent with the WLC model. In spite of the modest size
difference between the polymer length and the range of background correlations,
we found that, for sufficiently stiff test polymers, the
renormalized persistence length is uniquely determined in terms of the thermal
persistence length \(\ell_p\) and a ``disorder persistence length'' \(\ell_p^D\)
that characterizes the ambient disorder. On this basis, polymers of known
intrinsic stiffness can be used as generic quantitative probes of molecular
crowding. With our simple formula for the renormalized persistence length, their
tangent correlations and radial distribution are conveniently analyzed in terms
of the background disorder parameters.

It is an intriguing question whether our findings generalize to three dimensions.
We assume they do, at least for generic kinds of disorder such as a random distribution of spheres or bent rods: neither the concept of tilt symmetry nor the idea of random polymer-obstacle collisions are specific to two dimensions.
Our ``disorder persistence length'' has a simple definition and could in principle be measured using standard video microscopy techniques analyzed in the usual way (by measuring tangent-tangent correlations and radial distribution functions).
Therefore, we hope to inspire not only further numerical or analytical work on the matter, but also experimental studies under physiological conditions.
In this context, it would also be interesting to extend our analysis to the case of annealed disorder, which has recently been addressed experimentally for flexible polymers~\cite{benschuler2014}.

\section{Acknowledgements}

We thank Johannes Zierenberg, Niklas Fricke, Martin Marenz and Jan Kierfeld for fruitful discussions and beneficial advice and acknowledge financial support from the Leipzig Graduate School of Excellence GSC185 ``BuildMoNa'', from the German Science Foundation via FOR877 and SFB/TRR~102, the European Union and the Free State of Saxony, and from the Deutsch-Franz\"osische Hochschule (DFH-UFA).

%

\end{document}